\documentclass{cupconf}

  \checkfont{eurm10}
  \iffontfound
    \IfFileExists{upmath.sty}
      {\typeout{^^JFound AMS Euler Roman fonts on the system,
                   using the 'upmath' package.^^J}%
       \usepackage{upmath}}
      {\typeout{^^JFound AMS Euler Roman fonts on the system, but you
                   dont seem to have the}%
       \typeout{'upmath' package installed. cupconf.cls can take advantage
                 of these fonts,^^Jif you use 'upmath' package.^^J}%
      }
  \else
  \fi

  \checkfont{msam10}
  \iffontfound
    \IfFileExists{amssymb.sty}
      {\typeout{^^JFound AMS Symbol fonts on the system, using the
                'amssymb' package.^^J}%
       \usepackage{amssymb}%
         \let\leq=\leqslant
         \let\geq=\geqslant
      }{}
  \fi

  \IfFileExists{amsbsy.sty}
    {\typeout{^^JFound the 'amsbsy' package on the system, using it.^^J}%
     \usepackage{amsbsy}}
    {}

\usepackage{epsf}

\newcommand{\etal}{{et al.~}}

\newcommand{\gta}{\gtrsim}

\newcommand{\masyr}{\>{\rm mas}\,{\rm yr}^{-1}}
\newcommand{\kms}{\>{\rm km}\,{\rm s}^{-1}}
\newcommand{\pc}{\>{\rm pc}}
\newcommand{\kpc}{\>{\rm kpc}}

\newcommand{\Msun}{\>{\rm M_{\odot}}}

\newcommand{\Gyr}{\>{\rm Gyr}}

\newif\ifsubmode
\submodefalse

\newif\ifprintfig
\printfigtrue

\title[LMC Structure and Kinematics]{The Large Magellanic Cloud:\\
Structure and Kinematics}

\author[van der Marel]{Roeland P. van der Marel}

\affiliation{Space Telescope Science Institute, 3700 San Martin
Drive, Baltimore, MD 21218}

\begin{document}

\maketitle

\begin{abstract}
I review our understanding of the structure and kinematics of the
Large Magellanic Cloud (LMC), with a particular focus on recent
results. This is an important topic, given the status of the LMC as a
benchmark for studies of microlensing, tidal interactions, stellar
populations, and the extragalactic distance scale. I address the
observed morphology and kinematics of the LMC; the angles under which
we view the LMC disk; its in-plane and vertical structure; the LMC
self-lensing contribution to the total microlensing optical depth; the
LMC orbit around the Milky Way; and the origin and interpretation of
the Magellanic Stream. Our understanding of these topics is evolving
rapidly, in particular due to the many large photometric and kinematic
datasets that have become available in the last few years. It has now
been established that: the LMC is considerably elongated in its disk
plane; the LMC disk is thicker than previously believed; the LMC disk
may have warps and twists; the LMC may have a pressure-supported halo;
the inner regions of the LMC show unexpected complexities in their
vertical structure; and precession and nutation of the LMC disk plane
contribute measurably to the observed line-of-sight velocity
field. However, many open questions remain and more work is needed
before we can expect to converge on a fully coherent structural,
dynamical and evolutionary picture that explains all observed features
of the LMC.
\end{abstract}

% if document starts with a section,
% remove some space above using this command.

\firstsection 

\section{Introduction}
\label{s:intro}

The Large Magellanic Cloud (LMC) is one of our closest neighbor
galaxies at a distance of $\sim 50 \kpc$. The Sagitarrius dwarf is
closer at $\sim 24 \kpc$, but its contrast with respect to the Milky
Way foreground stars is so low that it was discovered only about a
decade ago. The LMC is therefore the closest, big, easily observable
galaxy from our vantage point in the Milky Way. As such, it has become
a benchmark for studies on various topics. It is of fundamental
importance for studies of stellar populations and the interstellar
medium (ISM), it is being used to study the presence of dark objects
in the Galactic Halo through microlensing (e.g., Alcock \etal 2000a),
and it plays a key role in determinations of the cosmological distance
scale (e.g., Freedman \etal 2001). For all these applications it is
important to have an understanding of the structure and kinematics of
the LMC. This is the topic of the present review. For information on
other aspects of the LMC, the reader is referred to the book by
Westerlund (1997). The book by van den Bergh (2000) discusses more
generally how the properties of the LMC compare to those of other
galaxies in the Local Group.

The Small Magellanic Cloud (SMC) at a distance of $\sim 62 \kpc$ is a
little further from us than the LMC, and is about 5 times less
massive. Its structure is more irregular than that of the LMC, and it
is less well studied and understood. Recent studies of SMC structure
and kinematics include the work by Hatzidimitriou \etal (1997),
Udalski \etal (1998), Stanimirovic \etal (1999, 2004), Kunkel, Demers
\& Irwin (2000), Cioni, Habing \& Israel (2000b), Zaritsky \etal
(2000, 2002), Crowl \etal (2001) and Maragoudaki \etal
(2001). However, our overall understanding of SMC structure and
kinematics has not evolved much since the reviews by Westerlund and
van den Bergh. The present review is therefore restricted to the LMC.

\section{Morphology}
\label{s:morphology}

The LMC consists of an outer body that appears elliptical in
projection on the sky, with a pronounced, off-center bar. The
appearance in the optical wavelength regime is dominated by the bar,
regions of strong star formation, and patchy dust absorption. The LMC
is generally considered an irregular galaxy as a result of these
characteristics. It is in fact the prototype of the class of galaxies
called ``Magellanic Irregulars'' (de Vaucouleurs \& Freeman
1973). Detailed studies of the morphological characteristics of the
LMC have been performed using many different tracers, including
optically detected starlight (Bothun \& Thompson 1988; Schmidt-Kaler
\& Gochermann 1992), stellar clusters (Lynga \& Westerlund 1963;
Kontizas \etal 1990), planetary nebulae (Meatheringham \etal 1988) and
non-thermal radio emission (Alvarez, Aparici \& May 1987). Recent
progress has come primarily from studies of stars on the Red Giant
Branch (RGB) and Asymptotic Giant Branch (AGB) and from studies of HI
gas.

%%% FIGURE 1 %%%

\begin{figure}
\null\bigskip
\epsfxsize=0.6\hsize
\centerline{\epsfbox{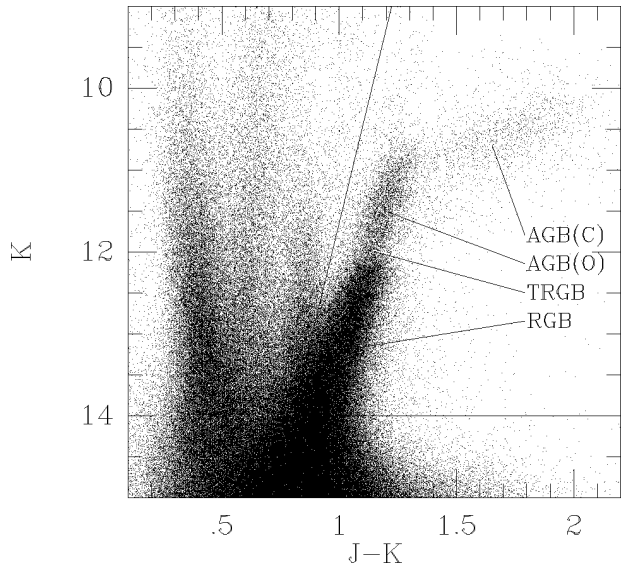}}
\caption{Near-IR $(J-K_s,K_s)$ CMD from 2MASS data for the LMC region
of the sky. Only a quarter of the data is shown, to avoid saturation
of the grey scale. The features due to the Red Giant Branch (RGB), the
Tip of the RGB (TRGB), the oxygen-rich AGB stars [AGB(O)], and the
carbon-rich AGB stars [AGB(C)] are indicated. The region enclosed by
the solid lines was used to extract stars for creation of the LMC
number density map in Figure~\ref{f:maps}.\label{f:cmds}}
\end{figure}

%%% END OF FIGURE 1 %%%

\subsection{Near-Infrared Morphology}
\label{ss:nearIR}

Recently, two important near-IR surveys have become available for
studies of the Magellanic Clouds, the Two Micron All Sky Survey
(2MASS; e.g., Skrutskie 1998) and the Deep Near-Infrared Southern Sky
Survey (DENIS; e.g., Epchtein \etal 1997). Cross-correlations of the
data from these surveys and from other catalogs are now available as
well (Delmotte \etal 2002). The surveys are perfect for a study of LMC
morphology and structure. Near-IR data is quite insensitive to dust
absorption, which is a major complicating factor in optical studies
(Zaritsky, Harris \& Thompson 1997; Zaritsky 1999; Udalski \etal 2000;
Alcock \etal 2000b). The surveys have superb statistics with of the
order of a million stars. Also, the observational strategy with three
near-IR bands ($J$, $H$ and $K_s$ in the 2MASS survey; $I$, $J$ and
$K_s$ in the DENIS survey) allows clear separation of different
stellar populations. In particular, the data are ideal for studies of
evolved RGB and AGB stars, which emit much of their light in the
near-IR. This is important for studies of LMC structure, because these
intermediate-age and old stars are more likely to trace the underlying
mass distribution of the LMC disk than younger populations that
dominate the light in optical images.

\begin{figure}
\null\bigskip
\epsfxsize=0.6\hsize
\centerline{\epsfbox{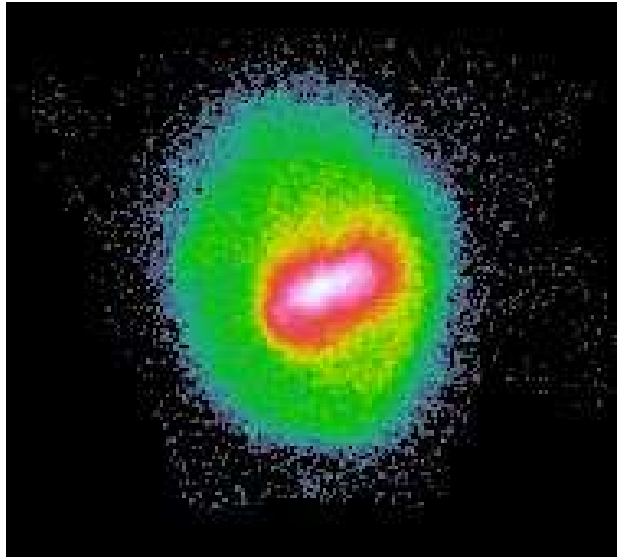}}
\caption{Surface number density distribution on the sky of RGB and AGB
stars in the LMC from van der Marel (2001). North is to the top and
east is to the left. The image covers an area of $23.55^{\circ} \times
21.55^{\circ}$. The Galactic foreground contribution was subtracted.
A color version of the image is available at
http://www.stsci.edu/$\sim$marel/lmc.html .\label{f:maps}}
\end{figure}

Figure~\ref{f:cmds} shows the $(J-K_s,K_s)$ color-magnitude diagram
(CMD) for the LMC region of the sky. Several finger-like features are
visible, each due to different stellar populations in the LMC or in
the foreground (Nikolaev \& Weinberg 2000; Cioni \etal 2000a; Marigo,
Girardi \& Chiosi 2003). The LMC features of primary interest in the
present context are indicated in the figure, namely the Red Giant
Branch (RGB), the Tip of the RGB (TRGB), the oxygen-rich AGB stars,
and the carbon-rich AGB stars (``carbon stars''). Van der Marel (2001)
used the color cut shown in the figure to extract a sample from the
2MASS and DENIS datasets that is dominated by RGB and AGB stars. These
stars were used to make the number-density map of the LMC shown in
Figure~\ref{f:maps}. Kontizas \etal (2001) showed the distribution on
the sky of $\sim 7000$ carbon stars identified by eye from optical
objective prism plates. Their map does not show all the rich detail
visible in Figure~\ref{f:maps}, but it is otherwise in good agreement
with it.
	
%%% FIGURE 2 %%%

%%% END OF FIGURE 2 %%%

The near-IR map of the LMC is surprisingly smooth. The morphology is
much less irregular than it is for the the younger stellar populations
that dominate the optical light. Apart from the central bar there is a
hint of some spiral structure, as discussed previously by, e.g., de
Vaucouleurs \& Freeman (1973). However, the spiral features all have
very low contrast with respect to their surroundings, and there is
certainly no well organized spiral pattern in the LMC. Quantitative
analysis can be performed on the basis of ellipse fits to the number
density contours. This yields a surface number density profile that
can be fit reasonably well fit by an exponential with a scale length
of $1.4^{\circ}$ ($1.3 \kpc$). The radial profiles of the ellipticity
$\epsilon$ (defined as $1 - q$, where $q$ is the axial ratio) and the
major axis position angle ${\rm PA}_{\rm maj}$ both show pronounced
variations as function of distance from the LMC center, due to the
presence of the central bar. However, at radii $r \gta 4^{\circ}$ the
contour shapes converge to an approximately constant position angle
${\rm PA}_{\rm maj} = 189.3^{\circ} \pm 1.4^{\circ}$ and ellipticity
$\epsilon = 0.199 \pm 0.008$.

%%% FIGURE 3 %%%

\begin{figure}
\null\bigskip
\epsfxsize=0.6\hsize
\centerline{\epsfbox{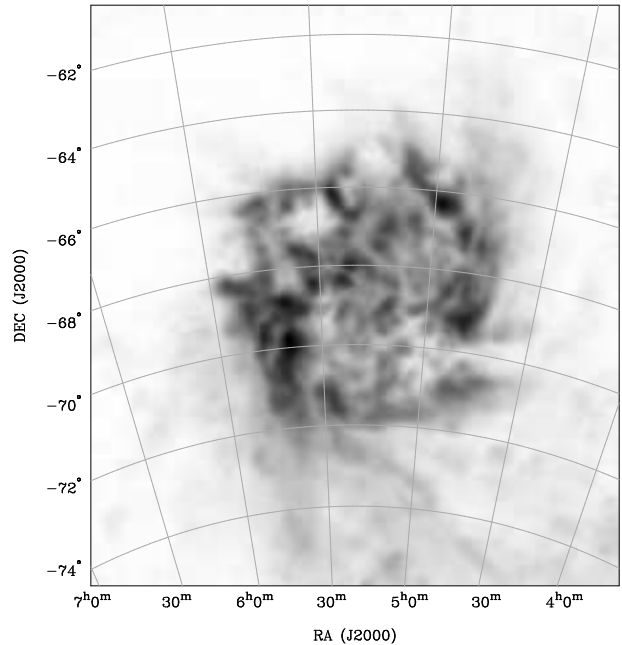}}
\caption{Peak brightness temperature image of HI in the LMC from
Staveley-Smith \etal (2003). The data are from a Parkes multibeam
survey and are sensitivity to spatial structure in the range $200
\pc$ to $10 \kpc$.  The orientation is the same as in
Figure~\ref{f:maps}, but the present image covers an area that is
approximately $2.5$ times smaller (i.e., $1.6$ in each
dimension).\label{f:HImap}}
\end{figure}

%%% END OF FIGURE 3 %%%

\subsection{HI Morphology}
\label{ss:HImorph}

To study the distribution of HI in the LMC on small scales requires
many pointings to cover the LMC at high angular resolution. Kim \etal
(1998) used the Australia Telescope Compact Array (ATCA) to obtain a
map that is sensitive on scales of 15--$500 \pc$. On these scales, the
morphology is dominated by HI filaments with numerous shells and
holes. The turbulent and fractal nature of the ISM on these scales is
the result of dynamical feedback into the ISM from star formation
processes. In the context of the present review we are more interested
in the large scale distribution of HI gas. This issue has been studied
for decades, including work by McGee \& Milton (1966), Rohlfs \etal
(1984) and Luks \& Rohlfs (1992). Most recently, Staveley-Smith \etal
(2003) obtained the map shown in Figure~\ref{f:HImap}, using a
multibeam survey with the Parkes telescope. This map is sensitive to
spatial structure in the range $200 \pc$ to $10 \kpc$. The HI
distribution is very patchy (see also Kim \etal 2003, who combined the
data of Kim \etal [1998] and Staveley-Smith \etal [2003] into a single
map that is sensitive to structures on all scales down to $15
\pc$). The HI in the LMC is not centrally concentrated and the
brightest areas are several degrees from the center. The overall
distribution is approximately circular and there is no sign of a
bar. All of these characteristics are strikingly different from the
stellar distribution shown in Figure~\ref{f:maps}.

\begin{figure}
\null\bigskip
\epsfxsize=0.6\hsize
\centerline{\epsfbox{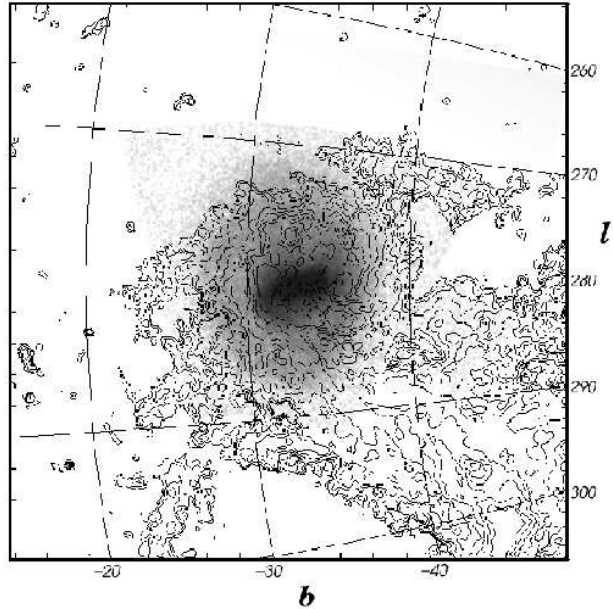}}
\caption{Contours of HI column density from Putman \etal (2003)
overlaid on the LMC near-IR star count map from
Figure~\ref{f:maps}. This figure is in Galactic coordinates, but
$(l,b)$ are shown so as to allow easy comparison with
Figures~\ref{f:maps} and~\ref{f:HImap}. North is $9^{\circ}$ clockwise
from the vertical direction on the page. The image covers an area that
is approximately $2.5$ times larger than Figure~\ref{f:maps} (i.e.,
$1.6$ times in each dimension). The figure was kindly provided by
M.~Putman.\label{f:HIoverlay}}
\end{figure}

Figure~\ref{f:HIoverlay} shows contours of the HI distribution on a
somewhat larger scale, from the HIPASS data presented in Putman \etal
(2003), overlaid on the LMC near-IR star count map from
Figure~\ref{f:maps}. This comparison suggests that collisionless
tracers such as RGB and AGB stars are best suited to study the
structure and mass distribution of the LMC disk, whereas HI gas may be
better suited to study the effects of tidal interactions. The stream
of gas towards the right of Figure~\ref{f:HIoverlay} is the start of
the Magellanic Bridge towards the SMC, which provides one of the many
pieces of evidence for strong tidal interactions between the LMC, the
Milky Way and the SMC.

%%% FIGURE 4 %%%

%%% END OF FIGURE 4 %%%

\section{Magellanic Stream}
\label{s:stream}

\begin{figure}
\null\bigskip
\epsfxsize=\hsize
\centerline{\epsfbox{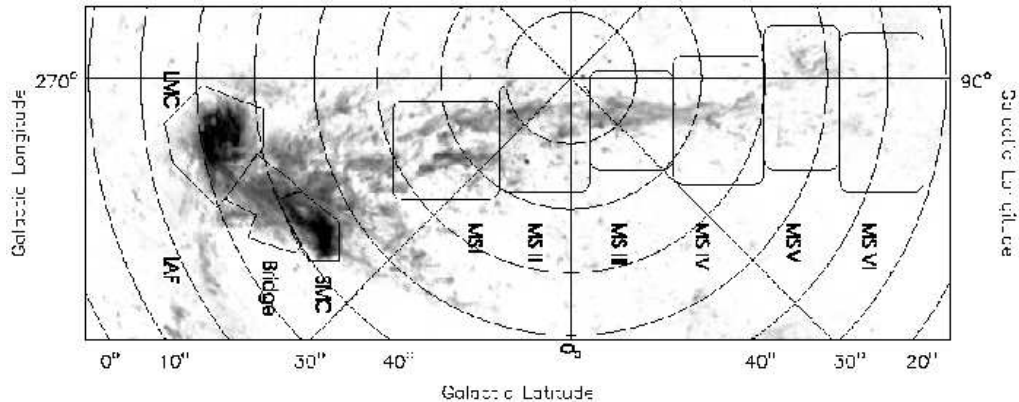}}
\caption{HI column density map of the Magellanic System in
galactic coordinates from Putman \etal (2003). The LMC and SMC are
indicated, as well as the Magellanic Bridge between them, several
individual gas clumps in the Magellanic Stream (MS\#), and the start of
the Leading Arm Feature (LAF). The orientation of the LMC on the page
is similar to that in Figure~\ref{f:HIoverlay}.\label{f:stream}}
\end{figure}

Figure~\ref{f:stream} shows a wide-area HI map of the Magallanic
System. The most prominent feature is the Magellanic Stream, a
$10^{\circ}$ wide filament with $\sim 2 \times 10^8 \Msun$ of neutral
hydrogen that spans more than $100^{\circ}$ across the sky (e.g.,
Westerlund 1997; Putman \etal 2003). It consists of gas that trails
the Magellanic Clouds as they orbit the Milky Way. A less prominent
leading gas component was recently discovered as well (Lu \etal 1998;
Putman \etal 1998), the start of which is seen at the left in
Figure~\ref{f:stream}.

%%% FIGURE 5 %%%

%%% END OF FIGURE 5 %%%

Many detailed theoretical models have been constructed for the
Magellanic Stream (e.g., Murai \& Fujimoto 1980; Lin \& Lynden-Bell
1982; Shuter 1992; Liu 1992; Heller \& Rohlfs 1994; Gardiner, Sawa \&
Fujimoto 1994; Moore \& Davis 1994; Lin, Jones \& Klemola 1995;
Gardiner \& Noguchi 1996; Yoshizawa \& Noguchi 2003; Mastropietro
\etal 2004; Conners \etal 2004). Models in which tidal stripping is
the dominating process have been particularly successful. The most
sophisticated recent calculations in this class are those by Yoshizawa
\& Noguchi (2003) and Conners \etal (2004), both of which build on
earlier work by Gardiner \& Noguchi (1996). In these models the LMC
and SMC form a gravitationally bound system that orbits the Milky
Way. The Magellanic Stream and the Leading Arm represent material that
was stripped from the SMC $\sim 1.5 \Gyr$ ago. This was the time of
the previous perigalactic passage, which coincided with a close
encounter between the Clouds. The models successfully reproduce many
properties of the Magellanic Stream, including its position,
morphology, width variation, and the velocity profile along the
Stream. The models also explain the presence of the Leading Arm, and
why it is less prominent than the trailing Stream.

Given the successes of tidal models, it has always been surprising that
no population of stars associated with the Stream has ever been found
(e.g., Irwin 1991; Guhathakurta \& Reitzel 1998; Majewski \etal
2003). In a tidal model where the stripping is dominated by gravity,
one might naively expect that both stars and gas are stripped equally.
However, galaxies generally have HI gas disks that are more extended
than the stellar distribution. Since material is preferentially stripped
from the outskirts of a galaxy, this can explain why there may not be
any stars associated with the Stream (Yoshizawa \& Noguchi 2003).
Alternatively, it has been argued that the lack of stars in the Stream
may point to important contributions from other physical processes
than tidal effects. For example, Moore \& Davis (1994) and
Mastropietro \etal (2004) suggest that the Stream consists of material
which was ram-pressure stripped from the LMC during its last passage
through a hot ($\sim 10^6$ K) ionized halo component of the Milky
Way. Another alternative was proposed by Heller \& Rohlfs (1994), who
suggested that hydrodynamical forces (rather than gravitational/tidal
forces) during a recent LMC-SMC interaction are responsible for the
existence of the Stream.

Models of the Magellanic Stream have traditionally used the properties
of the Stream to estimate the orbit of the LMC, rather than to base
the calculations on estimates of the LMC proper motion. An important
characteristic of the orbit is the present-day tangential velocity in
Galactocentric coordinates, for which values have been inferred that
include $v_{\rm LMC,tan} = 369 \kms$ (Lin \& Lynden-Bell 1982), $355
\kms$ (Shuter 1992), $352 \kms$ (Heller \& Rohlfs 1994), $339 \kms$
(Murai \& Fujimoto 1980), $320 \kms$ (Liu 1992) and $285 \kms$
(Gardiner \etal 1994; Gardiner \& Noguchi 1996), respectively. Proper
motion measurements have improved significantly over time, and it is
now known that $v_{\rm LMC,tan} = 281 \pm 41\kms$ (as discussed in
detail in Section~\ref{s:orbit}). This is consistent with the lower
range of the values predicted by models for the Stream. The data are
most consistent with the models of Gardiner \& Noguchi (1996) and
their follow-ups, which also provide some of the best fits to many
other properties of the Magellanic Stream. The observational error on
$v_{\rm LMC,tan}$ is almost small enough to start ruling out some of
the other models. It is possible that future proper motion
measurements may yield a much more accurate determination of the LMC
orbit. Models of the Magellanic Stream then hold the promise of
providing important constraints on the mass, shape, and radial density
profile of the Milky Way's dark halo (e.g., Lin \etal 1995).

\section{Orbit}
\label{s:orbit}

To understand the tidal interactions in the Magellanic System it is
important to know the orbit of the LMC around the Milky Way. This
requires knowledge of all three of the velocity components of the LMC
center of mass. The line-of-sight velocity can be accurately
determined from the Doppler velocities of tracers (see
Section~\ref{s:kinematics}). By contrast, determination of the
velocity in the plane of the sky is much more difficult. For the LMC,
proper motion determinations are available from the following sources:
Kroupa \etal (1994), using stars from the PPM Catalogue; Jones,
Klemola \& Lin (1994), using photographic plates with a 14 year epoch
span; Kroupa \& Bastian (1997), using Hipparcos data; Drake \etal
(2002), using data from the MACHO project; and Anguita, Loyola \&
Pedreros (2000) and Pedreros, Anguita \& Maza (2002), using CCD frames
with an 11 year epoch span. Some of the measurements pertain to fields
in the outer parts in the LMC disk, and require corrections for the
orientation and rotation of the LMC disk. The measurements are all
more-or-less consistent with each other to within the error bars. The
exception to this is the Anguita \etal result, which almost certainly
suffers from an unidentified systematic error. When this latter result
is ignored, the weighted average of the remaining measurements yields
proper motions towards the West and North of (van der Marel \etal
2002)
\begin{equation}
\label{weightB}
   \mu_W = -1.68 \pm 0.16 \masyr , \qquad \mu_N = 0.34 \pm 0.16 \masyr .
\end{equation}

Transformation of the proper motion to a space velocity in $\kms$
requires knowledge of the LMC distance $D_0$. Many techniques have
been used over the years to estimate the LMC distance, but
unfortunately, there continue to be systematic differences between the
results from different techniques that far exceed the formal
errors. It is beyond the scope of the present review to address this
topic in detail. Instead, the reader is refered to the recent reviews
by, e.g., Westerlund (1997), Gibson \etal (2000), Freedman \etal
(2001) and Alves (2004b). Freedman \etal adopt a distance modulus $m -
M = 18.50 \pm 0.10$ on the basis of a review of all published
work. This corresponds to $D_0 = 50.1 \pm 2.5 \kpc$. At this distance,
a proper motion of $1 \masyr$ corresponds to $238 \pm 12 \kms$ and 1
degree on the sky corresponds to $0.875 \pm 0.044 \kpc$.

Combination of the distance and proper motion of the LMC yields
velocities $v_W = -399 \kms$ and $v_N = 80 \kms$ towards the West and
North, respectively, with errors of $\sim 40 \kms$ in each
direction. This can be combined with the observed line-of-sight
velocity, $v_{\rm sys} = 262.2 \pm 3.4 \kms$
(see Section~\ref{ss:carbonstars}), to obtain the three-dimensional motion
of the LMC with respect to the Milky Way. It is usual to adopt a
Cartesian coordinate system $(X,Y,Z)$, with the origin at the Galactic
Center, the $Z$-axis pointing towards the Galactic North Pole, the
$X$-axis pointing in the direction from the sun to the Galactic
Center, and the $Y$-axis pointing in the direction of the sun's
Galactic Rotation. The observed LMC velocities must be corrected for
the reflex motion of the sun, which is easily done with use of
standard estimates for the position and velocity of the sun with
respect to the Galactic center. The $(X,Y,Z)$ position of the LMC is
then found to be (van der Marel \etal 2002)
\begin{equation}
\label{LMCposition}
  {\vec r}_{\rm LMC} = (-0.78,\> -41.55, \> -26.95) \kpc ,
\end{equation}
and its three-dimensional space velocity is
\begin{equation}
\label{LMCvelres}
  {\vec v}_{\rm LMC} = (  -56 \pm 36 , \> 
                          -219 \pm 23 , \>
                           186 \pm 35 ) \kms . 
\end{equation}
This corresponds to a distance of $49.53 \kpc$ from the Galactic
center, and a total velocity of $293 \pm 39 \kms$ in the
Galactocentric rest frame. The motion has a radial component of $84
\pm 7 \kms$ pointing away from the Galactic center, and a tangential
component of $281 \pm 41 \kms$. The proper motion of the SMC is known
only with errors of $\sim 0.8 \masyr$ in each coordinate (Kroupa \&
Bastian 1997), which is five times less accurate than for the LMC.
However, to within these errors the SMC is known to have a
galactocentric velocity vector that agrees with that of the LMC.

The combination of a small but positive radial velocity and a
tangential velocity that exceeds the circular velocity of the Milky
Way halo implies that the LMC must be just past pericenter in its
orbit. The calculation of an actual orbit requires knowledge of the
three-dimensional shape and the radial profile of the gravitational
potential of the Milky Way dark halo. Gardiner \etal (1994) and
Gardiner \& Noguchi (1996) calculated orbits in a spherical Milky Way
halo potential with a rotation curve that stays flat at $220 \kms$ out
to a galactocentric distance of at least $200 \kpc$. Such a potential
is consistent with our present understanding of the Milky Way dark
halo (Kochanek 1996; Wilkinson \& Evans 1999; Ibata \etal 2001). The
calculations properly take into account that the LMC and SMC orbit
each other while their center of mass orbits the Milky Way. However,
since the LMC is more massive than the SMC, its motion is not too
different from that of the LMC-SMC center of mass. For an assumed
present-day LMC velocity ${\vec v}_{\rm LMC}$ consistent with that
given in equation~(\ref{LMCvelres}) the apocenter to pericenter ratio
is inferred to be $\sim 2.5:1$. The perigalactic distance is $\sim 45
\kpc$ and the orbital period around the Milky Way is $\sim 1.5 \Gyr$.

\section{Kinematics}
\label{s:kinematics}

The kinematical properties of the LMC provide important clues to its
structure. Observations have therefore been obtained for many
tracers. The kinematics of gas in the LMC have been studied primarily
using HI (e.g., Rohlfs \etal 1984; Luks \& Rohlfs 1992; Kim \etal
1998). Discrete LMC tracers which have been studied kinematically
include star clusters (Freeman, Illingworth \& Oemler 1983; Schommer
\etal 1992), planetary nebulae (Meatheringham \etal 1988), HII regions
and supergiants (Feitzinger, Schmidt-Kaler \& Isserstedt 1977), and
carbon stars (Kunkel \etal 1997b; Graff \etal 2000; Alves \& Nelson
2000; van der Marel \etal 2002). A common result from all these
studies is that the line-of-sight velocity dispersion of the tracers
is generally at least a factor $\sim 2$ smaller than their rotation
velocity. This implies that the LMC is kinematically cold, and must
therefore to lowest approximation be a disk system.

\subsection{General Expressions}
\label{ss:generalexp}

\begin{figure}
\null\bigskip
\epsfxsize=0.6\hsize
\centerline{\epsfbox{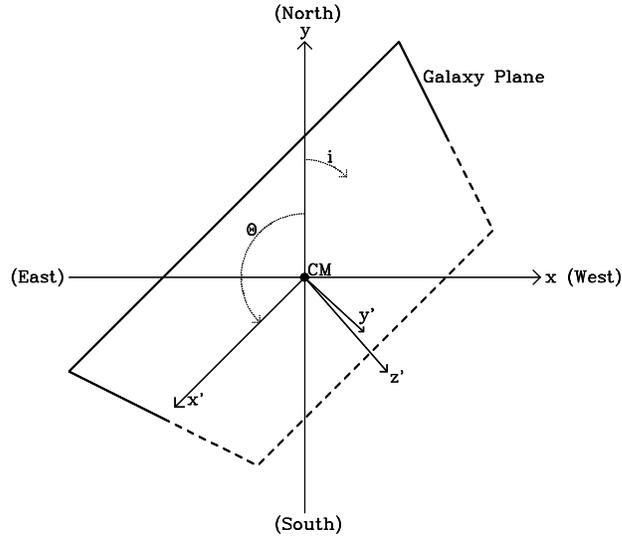}}
\caption{Schematic illustration of the observer's view of the LMC disk. The
plane of the disk is titled diagonally out of the paper. The
inclination $i$ is the angle between the $(x,y)$ plane of the sky, and
the $(x',y')$ plane of the galaxy disk. The $x'$-axis is the line of
nodes, defined as the intersection of the $(x,y)$ plane of the sky and
the $(x',y')$ plane of the galaxy disk. The angle $\Theta$ is the
position angle of the line of nodes in the plane of the sky.
\label{f:drawviewang}}
\end{figure}

To understand the kinematics of an LMC tracer population it is
necessary to have a general model for the line-of-sight velocity field
that can be fit to the data. All studies thus far have been based on
the assumption that the mean streaming (i.e., the rotation) in the
disk plane can be approximated to be circular. However, even with this
simplifying assumption it is not straightforward to model the
kinematics of the LMC, because it is so near to us. Its main body
spans more than $20^{\circ}$ on the sky and one therefore cannot make
the usual approximation that ``the sky is flat'' over the area of the
galaxy. Spherical trigonometry must be used, which yields the general
expression (van der Marel \etal 2002):
\begin{eqnarray}
\label{vfield}
   v_{\rm los}(\rho,\Phi) 
      &=& s \, V(R') f \sin i \cos (\Phi-\Theta) \nonumber \\
      &+& v_{\rm sys} \cos \rho \nonumber \\
      &+& v_t \sin \rho \cos (\Phi -\Theta_t) \nonumber \\
      &+& D_0 (di/dt) \sin \rho \sin (\Phi-\Theta) , 
\end{eqnarray}
with 
\begin{equation}
\label{Rfdef}
  R' = D_0 \sin \rho / f , \qquad 
  f        \equiv { {\cos i \cos \rho -
                     \sin i \sin \rho \sin (\Phi-\Theta)} \over 
                    { {[\cos^2 i \cos^2 (\Phi-\Theta) + 
                                 \sin^2 (\Phi-\Theta)]^{1/2}} } } .
\end{equation}
In this equation, $v_{\rm los}$ is the observed component of the
velocity along the line of sight. The quantities $(\rho,\Phi)$
identify the position on the sky: $\rho$ is the angular distance from
the center and $\Phi$ is the position angle with respect to the center
(measured from North over East). The kinematical center is at the
center of mass (CM) of the galaxy. The quantities $(v_{\rm sys}, v_t,
\Theta_t)$ describe the velocity of the CM in an inertial frame in
which the sun is at rest: $v_{\rm sys}$ is the systemic velocity along
the line of sight, $v_t$ is the transverse velocity, and $\Theta_t$ is
the position angle of the transverse velocity on the sky. The angles
$(i,\Theta)$ describe the direction from which the plane of the galaxy
is viewed: $i$ is the inclination angle ($i=0$ for a face-on disk),
and $\Theta$ is the position angle of the line of nodes, as
illustrated in Figure~\ref{f:drawviewang}. The line-of-nodes is the
intersection of the galaxy plane and the sky plane. The velocity
$V(R')$ is the rotation velocity at cylindrical radius $R'$ in the
disk plane. $D_0$ is the distance to the CM, and $f$ is a geometrical
factor. The quantity $s = \pm 1$ is the `spin sign' that determines in
which of the two possible directions the disk rotates.

%%% FIGURE 6 %%%

%%% END OF FIGURE 6 %%%

The first term in equation~(\ref{vfield}) corresponds to the internal
rotation of the LMC. The second term is the part of the line-of-sight
velocity of the CM that is seen along the line of sight, and the third
term is the part of the transverse velocity of the CM that is seen
along the line of sight. For a galaxy that spans a small area on the
sky (very small $\rho$), the second term is simply $v_{\rm sys}$ and
the third term is zero. However, the LMC does not have a small angular
extent and the inclusion of the third term is particularly
important. It corresponds to a solid-body rotation component. Given
the LMC transverse velocity implied by equation~(\ref{weightB}), it
rises to an amplitude of $71 \kms$ at $\rho = 10^{\circ}$, which
significantly exceeds the amplitude of the intrinsic rotation
contribution (first term of eq.~[\ref{vfield}]) at that radius. The
fourth term in equation~(\ref{vfield}) describes the line-of-sight
component due to changes in the inclination of the disk with time, as
are expected due to precession and nutation of the LMC disk plane as
it orbits the Milky Way (Weinberg 2000). This term also corresponds to
a solid-body rotation component.

The general expression in equation~(\ref{vfield}) appears complicated,
but it is possible to gain some intuitive insight by considering some
special cases. Along the line of nodes one has that $\sin
(\Phi-\Theta) = 0$ and $\cos (\Phi-\Theta) = \pm 1$, so that
\begin{equation}
\label{vlosalong}
   {\hat v}_{\rm los} ({\rm along}) = 
          \pm [v_{tc} \sin \rho - 
               V(D_0 \tan \rho) \sin i \cos \rho] .
\end{equation}
Here it has been defined that ${\hat v}_{\rm los} \equiv v_{\rm los} -
v_{\rm sys} \cos \rho \approx v_{\rm los} - v_{\rm sys}$. The quantity
$v_{tc} \equiv v_t \cos (\Theta_t - \Theta)$ is the component of the
transverse velocity vector in the plane of the sky that lies along the
line of nodes; similarly, $v_{ts} \equiv v_t \sin (\Theta_t - \Theta)$
is the component perpendicular to the line of nodes. Perpendicular to
the line of nodes one has that $\cos (\Phi-\Theta) = 0$ and $\sin
(\Phi-\Theta) = \pm 1$, and therefore
\begin{equation}
\label{vlosperp}
   {\hat v}_{\rm los} ({\rm perpendicular}) = \pm w_{ts} \sin \rho .
\end{equation}
Here it has been defined that $w_{ts} = v_{ts} + D_0 (di/dt)$. This
implies that perpendicular to the line of nodes ${\hat v}_{\rm los}$
is linearly proportional to $\sin \rho$. By contrast, along the line
of nodes this is true only if $V(R')$ is a linear function of
$R'$. This is not expected to be the case, because galaxies do not
generally have solid-body rotation curves; disk galaxies tend to have
flat rotation curves, at least outside the very center. This implies
that, at least in principle, both the position angle $\Theta$ of the
line of nodes and the quantity $w_{ts}$ are uniquely determined by the
observed velocity field: $\Theta$ is the angle along which the
observed ${\hat v}_{\rm los}$ are best fit by a linear proportionality
with $\sin \rho$, and $w_{ts}$ is the proportionality constant.

%%% FIGURE 7 %%%

\begin{figure}
\null\bigskip
\epsfxsize=0.8\hsize
\centerline{\epsfbox{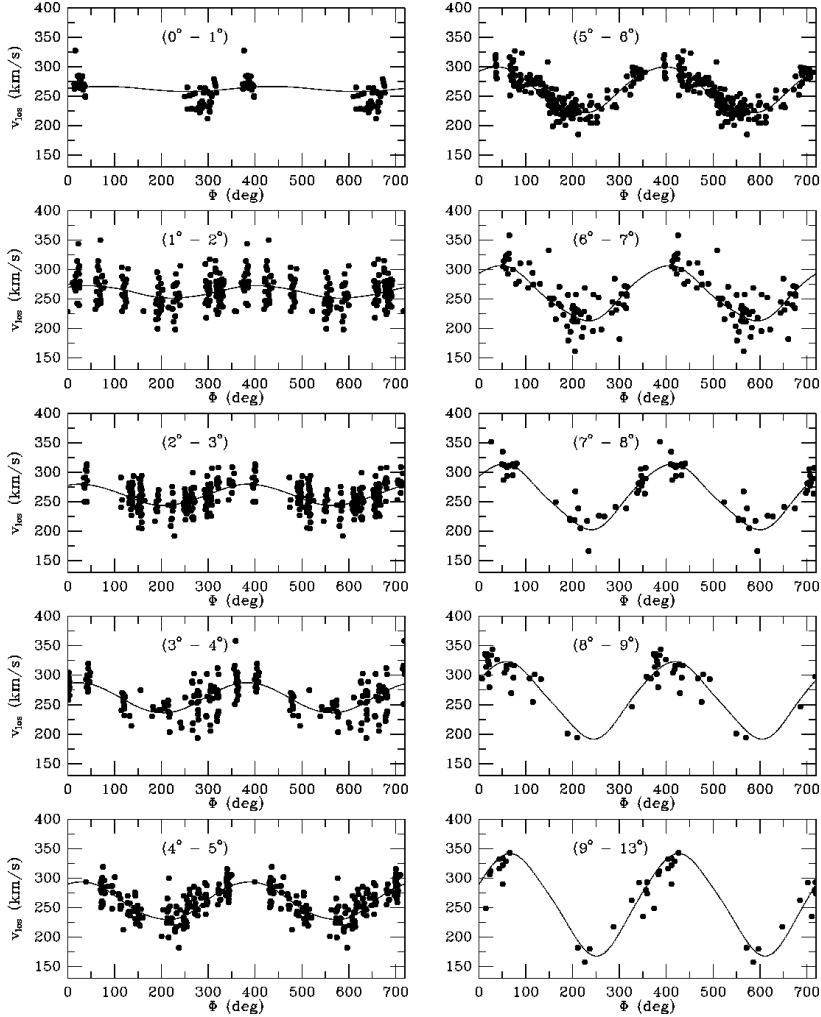}}
\caption{Carbon star line-of-sight velocity data from Kunkel, Irwin \&
Demers (1997a) and Hardy \etal (unpublished), as function of position
angle $\Phi$ on the sky. The displayed range of the angle $\Phi$ is
$0^{\circ}$--$720^{\circ}$, so each star is plotted twice. Each panel
corresponds to a different range of angular distances $\rho$ from the
LMC center, as indicated. The curves show the predictions of the
best-fitting circularly-rotating disk model from van der Marel \etal
(2002).\label{f:fit}}
\end{figure}

%%% END OF FIGURE 7 %%%

\subsection{Carbon Star Kinematics}
\label{ss:carbonstars}

Among the discrete tracers in the LMC that have been studied
kinematically, carbon stars have yielded some of the largest and most
useful datasets in recent years. Van der Marel \etal (2002) fitted the
general velocity field expression in equation~(\ref{vfield}) to the
data for 1041 carbon stars, obtained from the work of Kunkel \etal
(1997a) and Hardy, Schommer \& Suntzeff (unpublished). The combined
dataset samples both the inner and the outer parts of the LMC,
although with a discontinuous distribution in radius and position
angle. Figure~\ref{f:fit} shows the data, with the best model fit
overplotted.  Overall, the model provides a good fit to the data.

%%% FIGURE 8 %%%

\begin{figure}
\null\bigskip
\epsfxsize=0.6\hsize
\centerline{\epsfbox{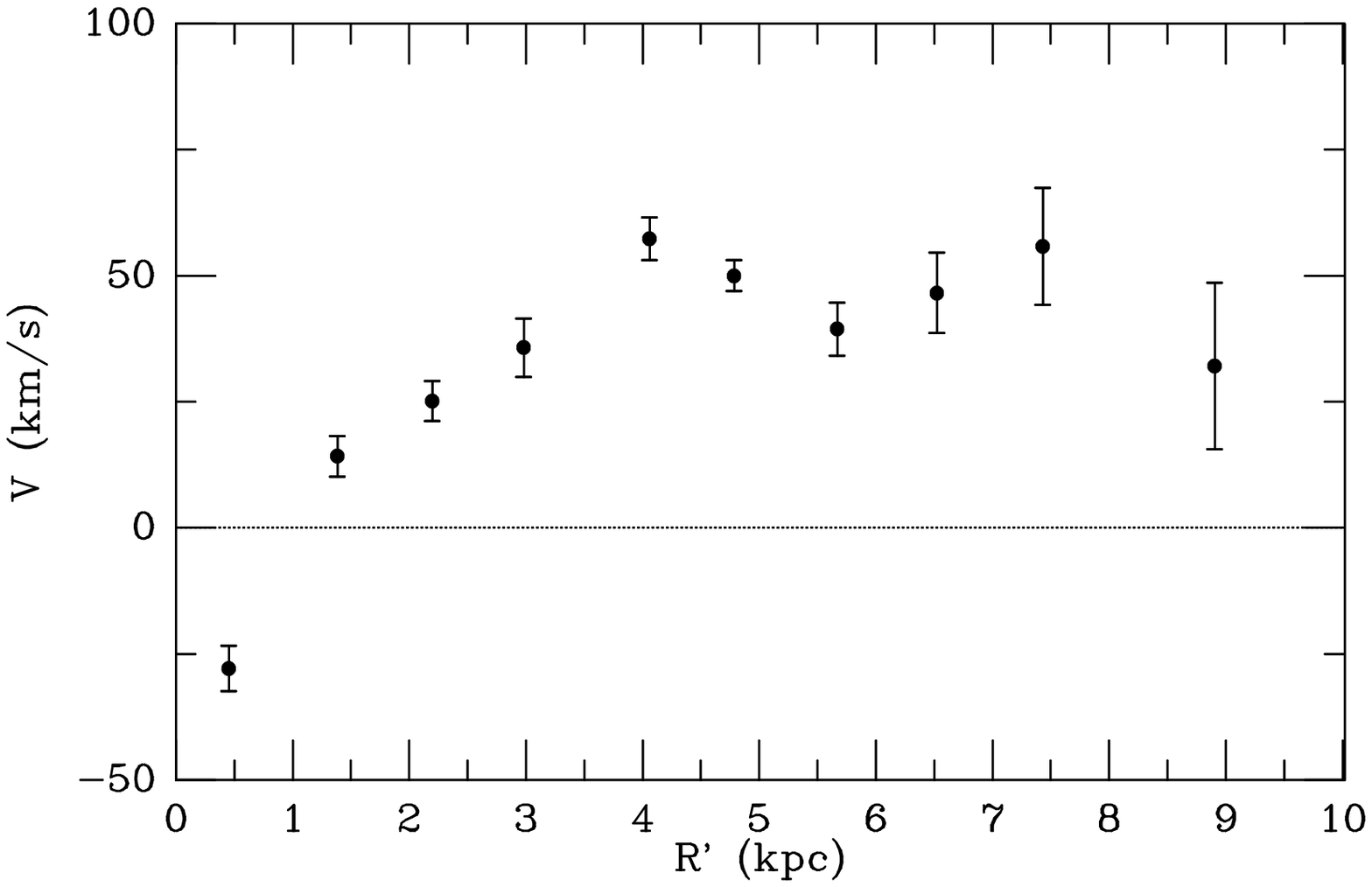}}
\caption{Rotation velocity $V$ of carbon stars in the plane of the LMC
disk as function of the cylindrical radius $R'$ in kpc from the carbon
star fits of van der Marel \etal (2002). For reference, the
exponential disk scale length of the LMC is approximately $1.4
\kpc$.\label{f:rotcurve}}
\end{figure}

%%% END OF FIGURE 8 %%%

As discussed in Section~\ref{ss:generalexp}, the line-of-nodes
position angle is uniquely determined by the data; the model yields
$\Theta = 129.9^{\circ} \pm 6.0^{\circ}$. The LMC inclination cannot
be determined kinematically, but it is known reasonably well from
other considerations (see Section~\ref{s:viewing}). With both viewing
angles and the LMC proper motion known, the rotation curve $V(R')$
follows from equation~(\ref{vlosalong}). The result is shown in
Figure~\ref{f:rotcurve}. The inferred rotation curve $V(R')$ rises
linearly in the central region and is roughly flat at $V \approx 50
\kms$ for $R' \gta 4 \kpc$. The negative value at the innermost radius
$R' \approx 0.5 \kpc$ has limited significance; it is probably
affected by the sparse coverage of the data at these radii (see
Figure~\ref{f:fit}), as well as potential non-circular streaming
motions in the region of the bar. The error bars on $V(R')$ in
Figure~\ref{f:rotcurve} reflect the random errors due to the sampling
of the data. However, there are also contributions from the errors in
the line-of-nodes position angle $\Theta$, the LMC proper motion and
the inclination $i$. In particular, $V(R')$ scales as $1/\sin i$;
and if the component $v_{tc}$ of the transverse velocity vector along
the line of nodes is larger than given in equation~(\ref{weightB}),
then $V(R')$ will go up. When all uncertainties are properly accounted
for, the amplitude of the flat part of the rotation curve and its
formal error become $V = 49.8 \pm 15.9 \kms$.

When asymmetric drift is corrected for, the circular velocity in the
disk plane can be calculated to be $V_{\rm circ} = 64.8 \pm 15.9
\kms$.  The implied total mass of the LMC inside the last measured
data point is therefore $M_{\rm LMC} (8.9 \kpc) = (8.7 \pm 4.3) \times
10^9 \Msun$. By contrast, the total stellar mass of the LMC disk is
$\sim 2.7 \times 10^9 \Msun$ and the mass of the neutral gas in the
LMC is $\sim 0.5 \times 10^9 \Msun$ (Kim \etal 1998). The combined
mass of the visible material in the LMC is therefore insufficient to
explain the dynamically inferred mass. The LMC must therefore be
embedded in a dark halo. This is consistent with the fact that the
observed rotation curve amplitude is relatively flat as a function of
radius. The LMC tidal radius can be calculated to be $r_t = 15.0 \pm
4.5 \kpc$, which corresponds to an angle on the sky of $17.1^{\circ}
\pm 5.1^{\circ}$. The uncertainty in the tidal radius is due primarily
to our ignorance of how far the LMC dark halo extends. Either way, the
tidal radius extends beyond the region for which most observations of
the main body of the LMC are available. However, it should be kept in
mind that the tidal radius marks the position beyond which material
becomes unbound. The structure of a galaxy can be altered well inside
of this radius (Weinberg 2000).

As discussed in Section~\ref{ss:generalexp}, the line-of-sight
velocity field constrains the value of $w_{ts} = v_{ts} + D_0
(di/dt)$. The carbon stars yield $w_{ts} = -402.9 \pm 13.0
\kms$. Given our knowledge of the proper motion and distance of the
LMC (see Section~\ref{s:orbit}) this implies that $di/dt = -0.37 \pm 0.22
\masyr = -103 \pm 61 \>\> {\rm degrees}/\Gyr$. The LMC is the first
galaxy for which it has been possible to measure $di/dt$. The $N$-body
simulations by Weinberg (2000) show that the tidal torques from the
Milky Way are expected to induce precession and nutation in the
symmetry axis of the LMC disk plane. Although a detailed data-model
comparison is not possible at the present time, it is comforting to
note that the observed $di/dt$ is of the same order as the rate of
change of the disk orientation seen in the simulations.

\subsection{HI Kinematics}
\label{ss:HIkin}

HI gas provides another powerful method to study the kinematics of the
LMC. High quality data is available from, e.g., Kim \etal (1998).
Unfortunately, the kinematical analysis presented by Kim \etal was not
as general as that discussed above for carbon stars. They did not
leave $w_{ts}$ as a free parameter in the fit. Instead, they corrected
their data at the outset for the transverse motion of the LMC using
the proper motion measured by Jones \etal (1994), $(\mu_W,\mu_N) =
(-1.37 \pm 0.28, -0.18 \pm 0.27) \masyr$, and assumed that $di/dt =
0$. This fixes $w_{ts} = -175 \pm 72 \kms$, which is inconsistent with
the value inferred from the carbon star velocity field. Kim \etal
obtained a kinematic line of nodes that is both twisting with radius
and inconsistent with the value determined from the carbon stars. In
addition, they inferred a rotation curve for which the amplitude
exceeds that in Figure~\ref{f:rotcurve} by $\sim 40\%$. It is likely
that these results are affected by the imposed value of $w_{ts}$. A
more general analysis of the HI kinematics is therefore desirable, but
unfortunately, is not currently available. The same limitations apply
to many of the other published studies of LMC tracer kinematics cited
at the start of Section~\ref{s:kinematics}.

Independent of how the data are analyzed, it is likely that
collisionless tracers provide a more appropriate means to study the
structure of the LMC using equilibrium models than does HI gas. The
dynamical center of the carbon star velocity field is found to be
consistent (to within the $\sim 0.4^{\circ}$ per-coordinate errors)
with both the center of the bar and the center of the outer stellar
contours. However, it has long been known that the center of the HI
rotation velocity field does not coincide with the center of the bar,
and it also doesn't coincide with the center of the outer contours of
the stellar distribution. It is offset from both by $\sim 1
\kpc$. This indicates that the HI gas may not be in equilibrium in the
the gravitational potential. Added evidence for this comes from the
fact that the LMC and the SMC are enshrouded in a common HI envelope,
and that they are connected by a bridge of HI gas (see
Figure~\ref{f:stream}). Even at small radii the LMC gas disk appears
to be subject to tidal disturbances (see Figure~\ref{f:HIoverlay})
that may well affect the velocity field.

\section{Viewing Angles}
\label{s:viewing}

A disk that is intrinsically circular will appear elliptical in
projection on the sky. The viewing angles of the disk (see
Figure~\ref{f:drawviewang}) are then easily determined: the
inclination is $i = \arccos (1-\epsilon)$, where $\epsilon$ is the
apparent ellipticity on the sky, and the line-of-nodes position angle
$\Theta$ is equal to the major axis position angle ${\rm PA}_{\rm
maj}$ of the projected body. The viewing angles of the LMC have often
been estimated under this assumption, using the projected contours for
many different types of tracers (de Vaucouleurs \& Freeman 1973;
Bothun \& Thompson 1988; Schmidt-Kaler \& Gochermann 1992; Weinberg \&
Nikolaev 2001; Lynga \& Westerlund 1963; Kontizas \etal 1990;
Feitzinger \etal 1977; Kim \etal 1998; Alvarez \etal 1987). However,
it now appears that this was incorrect.  The kinematics of carbon
stars imply $\Theta = 129.9^{\circ} \pm 6.0^{\circ}$ (see
Section~\ref{ss:carbonstars}), whereas the near-IR morphology of the
LMC implies ${\rm PA}_{\rm maj} = 189.3^{\circ} \pm 1.4^{\circ}$ (see
Section~\ref{s:morphology}). The result that $\Theta \not= {\rm
PA}_{\rm maj}$ implies that the LMC cannot be intrinsically
circular. The value of ${\rm PA}_{\rm maj}$ is quite robust; studies
of other tracers have yielded very similar results, although often
with larger error bars. The result that $\Theta \not= {\rm PA}_{\rm
maj}$ therefore hinges primarily on our confidence in the inferred
value of $\Theta$. There have been other kinematical studies of the
line of nodes, in addition to that described in
Section~\ref{ss:carbonstars}. These have generally yielded values of
$\Theta$ that are both larger and twisting with radius (e.g., Kim
\etal 1998; Alves \& Nelson 2000). However, the accuracy of these
results is suspect because of the important simplifying assumptions
that were made in the analyses (see Section~\ref{ss:HIkin}). No
allowance was made for a potential solid-body rotation component in
the velocity field due to precession and nutation of the LMC disk,
which is both predicted theoretically (Weinberg 2000) and implied
observationally by the carbon star data (see
Section~\ref{ss:carbonstars}).

Arguably the most robust way to determine the LMC viewing angles is to
use geometrical considerations, rather than kinematical ones. For an
inclined disk, one side will be closer to us than the other. Tracers
on that one side will appear brighter than similar tracers on the
other side.  This method does not rely on absolute distances or
magnitudes, which are notoriously difficult to estimate, but only on
relative distances or magnitudes. To lowest order, the difference in
magnitude between a tracer at the galaxy center and a similar tracer
at a position $(\rho,\Phi)$ in the disk (as defined in
Section~\ref{ss:generalexp}) is
\begin{equation}
\mu = 
   \Big ( { {5\,\pi} \over {180\ln 10} } \Big ) \> 
     \rho \tan i \sin(\Phi-\Theta) ,
\label{dmagtaylor}
\end{equation}
where the angular distance $\rho$ is expressed in degrees. The
constant in the equation is $(5\pi) / (180\ln 10) = 0.038$
magnitudes. Hence, when following a circle on the sky around the
galaxy center one expects a sinusoidal variation in the magnitudes of
tracers. The amplitude and phase of the variation yield estimates of
the viewing angles $(i,\Theta)$.

%%% FIGURE 9 %%%

\begin{figure}
\null\bigskip
\epsfxsize=0.5\hsize
\centerline{\epsfbox{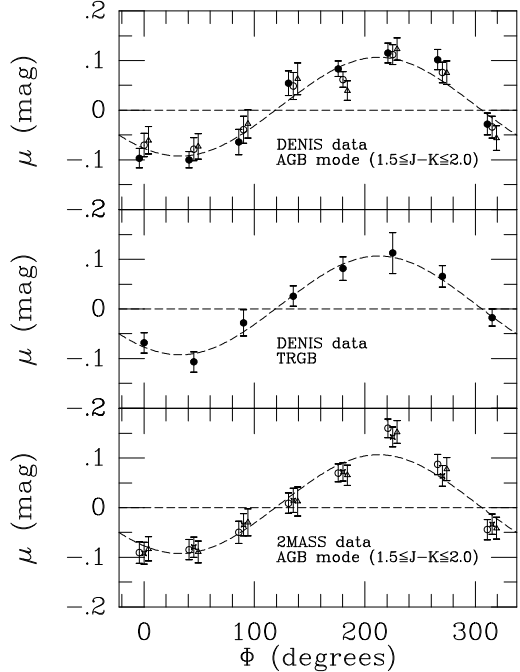}}
\caption{Variations in the magnitude of tracers as function of
position angle $\Phi$ from van der Marel \& Cioni (2001). {\it (a)}
Modal magnitude of AGB stars in the DENIS data with colors $1.5 \leq
J-K_s \leq 2.0$. {\it (b)} TRGB magnitude from DENIS data. {\it (c)}
As (a), but using data from the 2MASS Point Source Catalog. All panels
refer to an annulus of radius $2.5^{\circ} \leq \rho \leq 6.7^{\circ}$
around the LMC center.  Filled circles, open circles, four-pointed
stars and open triangles refer to the $I$, $J$, $H$ and $K_s$-band,
respectively. Results in different bands are plotted with small
horizontal offsets to avoid confusion. The dashed curve shows the
predictions for an inclined disk with viewing angles $i =
34.7^{\circ}$ and $\Theta = 122.5^{\circ}$.\label{f:banddep}}
\end{figure}

%%% END OF FIGURE 9 %%%

Van der Marel \& Cioni (2001) used a polar grid on the sky to divide
the LMC area into several rings, each consisting of a number of
azimuthal segments. The data from the DENIS and 2MASS surveys were
used for each segment to construct near-IR CMDs similar to that shown
in Figure~\ref{f:cmds}. For each segment the modal magnitude
(magnitude where the luminosity function peaks) was determined for
carbon-rich AGB stars selected by color, as had been suggested by
Weinberg \& Nikolaev (2001). Figure~\ref{f:banddep} shows the inferred
variation in magnitude as function of position angle $\Phi$ for the
radial range $2.5^{\circ} \leq \rho \leq 6.7^{\circ}$. The expected
sinusoidal variations are confidently detected. The top panel shows
the results for stars selected from the DENIS survey with the color
selection criterion $1.5 \leq J-K \leq 2.0$. The bottom panel shows
the results from the 2MASS survey with the same color selection. The
same sinusoidal variations are seen, indicating that there are no
relative calibration problems between the surveys. Also, the same
variations are seen in the $I$, $J$, $H$ and $K_s$ bands, which
implies that the results are not influenced significantly by dust
absorption. The middle panel shows the variations in the TRGB
magnitudes as a function of position angle, from the DENIS data. RGB
stars show the same variations as the AGB stars, suggesting that the
results are not influenced significantly by potential peculiarities
associated with either of these stellar populations. The observed
variations can therefore be confidently interpreted as a purely
geometrical effect. The implied viewing angles are $i = 34.7^{\circ}
\pm 6.2^{\circ}$ and $\Theta = 122.5^{\circ} \pm 8.3^{\circ}$. The
$\Theta$ value thus inferred geometrically is entirely consistent with
the value inferred kinematically
(see Section~\ref{ss:carbonstars}). Moreover, there is an observed drift
in the center of the LMC isophotes at large radii which is consistent
with both estimates, when interpreted as a result of viewing
perspective (van der Marel 2001).

The aforementioned analyses are sensitive primarily to the structure
of the outer parts of the LMC. Several other studies of the viewing
angles have focused mostly on the region of the bar, which samples
only the central few degrees. Many of these studies have been based on
Cepheids. Their period-luminosity relation allows calculation of the
distance to each individual Cepheid from a light curve. The relative
distances of the Cepheids in the sample can then be analyzed in
similar fashion as discussed above to yield the LMC viewing
angles. Cepheid studies in the 1980s didn't have many stars to work
with.  Caldwell \& Coulson (1986) analyzed optical data for 73
Cepheids and obtained $i = 29^{\circ} \pm 6^{\circ}$ and $\Theta =
142^{\circ} \pm 8^{\circ}$. Laney \& Stobie (1986) obtained $i =
45^{\circ} \pm 7^{\circ}$ and $\Theta = 145^{\circ} \pm 17^{\circ}$
from 14 Cepheids, and Welch \etal (1987) obtained $i = 37^{\circ} \pm
16^{\circ}$ and $\Theta = 167^{\circ} \pm 42^{\circ}$ from 23
Cepheids, both using near-IR data. The early Cepheid studies have now
all been superseded by the work of Nikolaev \etal (2004). They
analyzed a sample of more than 2000 Cepheids with lightcurves from
MACHO data. Through use of photometry in five different bands,
including optical MACHO data and near-IR 2MASS data, each star could
be individually corrected for dust extinction. From a planar fit to
the data they obtained $i = 30.7^{\circ} \pm 1.1^{\circ}$ and $\Theta
= 151.0^{\circ} \pm 2.4^{\circ}$. Other recent work has used the
magnitude of the Red Clump to analyze the relative distances of
different parts of the LMC. Olsen \& Salyk (2002) obtained $i =
35.8^{\circ} \pm 2.4^{\circ}$ and $\Theta = 145^{\circ} \pm
4^{\circ}$, also from an analysis that was restricted mostly to the
the inner parts of the LMC.

There is one caveat associated with all viewing angle results for the
central few degrees of the LMC. Namely, it appears that the stars in
this region are not distributed symmetrically around a single
well-defined plane, as discussed in detail in
Section~\ref{ss:nonplanar}. In the present context we are mainly
concerned with the influence of this on the inferred viewing
angles. Olsen \& Salyk (2002) perform their viewing angle fit by
ignoring fields south-west of the bar, which do not seem to agree with
the planar solution implied by their remaining fields. By contrast,
Nikolaev \etal (2004) fit all the stars in their sample, independent
of whether or not they appear to be part of the main disk
plane. Clearly, the $(i,\Theta)$ results of Olsen \& Salyk and
Nikolaev \etal are the best-fitting parameters of well-posed
problems. However, it is somewhat unclear whether they can be assumed
to be unbiased estimates of the actual LMC viewing angles. For a
proper understanding of this issue one would need to have both an empirical
and a dynamical understanding of the nature of the extra-planar
structures in the central region of the LMC. Only then is it possible
to decide whether the concept of a single disk plane is at all
meaningful in this region, and which data should be included or
excluded in determining its parameters. This is probably not an issue
for the outer parts of the LMC, given that the AGB star results of van
der Marel \& Cioni (2001) provide no evidence for extra-planar
structures at radii $\rho \geq 2.5^{\circ}$.

In summary, all studies agree that $i$ is in the range
$30^{\circ}$--$40^{\circ}$. At large radii, $\Theta$ appears to be in
the range $115^{\circ}$--$135^{\circ}$. By contrast, at small radii
all studies indicate that $\Theta$ is in the range
$140^{\circ}$--$155^{\circ}$. As mentioned, it is possible that the
results at small radii are systematically in error due to the presence
of out-of-plane structures. Alternatively, it is quite well possible
that there are true radial variations in the LMC viewing angles due to
warps and twists of the disk plane. Many authors have suggested this
as a plausible interpretation of various features seen in LMC datasets
(van der Marel \& Cioni 2001; Olsen \& Salyk 2002; Subramaniam 2003;
Nikolaev \etal 2004). Moreover, numerical simulations have shown that
Milky Way tidal effects can drive strong warps in the LMC disk plane
(Mastropietro \etal 2004).

\section{Ellipticity}
\label{s:ellip}

The inferred LMC viewing angles can be used to deproject the observed
morphology that is seen in projection on the sky (van der Marel
2001). This yields an in-plane ellipticity $\epsilon$ in the range
$\sim 0.2$--$0.3$, depending somewhat on the adopted viewing angles;
e.g., the Nikolaev et al.~(2004) angles give $\epsilon = 0.21$ and the
van der Marel \& Cioni (2001) angles give $\epsilon = 0.31$. The
conclusion that the LMC is elongated is in itself not surprising. The
dark matter halos predicted by cosmological simulations are generally
triaxial (e.g., Dubinski \& Carlberg 1991), and the gravitational
potential in the equatorial plane of such halos does not have circular
symmetry. So it is generally expected that disk galaxies are elongated
rather than circular. Furthermore, it is possible to construct
self-consistent dynamical models for elliptical disks (e.g., Teuben
1987). What is surprising is that the LMC ellipticity is fairly
large. Studies of the apparent axis ratio distribution of spiral
galaxy disks (Binney \& de Vaucouleurs 1981; Lambas, Maddox \& Loveday
1992) of the structure of individual spiral galaxies (Rix \& Zaritsky
1995; Schoenmakers, Franx \& de Zeeuw 1997; Kornreich, Haynes \&
Lovelace 1998; Andersen \etal 2001) and of the scatter in the
Tully-Fisher relation (Franx \& de Zeeuw 1992) indicate that the
average (deprojected) ellipticity of spiral galaxies is only
5--10\%. So while spiral galaxies are generally elongated, their
elongation is usually smaller than inferred here for the LMC. Of
course, galaxies of type Sm and Im are (by definition) more irregular
and lopsided than spirals. So it is not a priori clear whether or not
the LMC is atypically elongated for its Hubble type.

It is interesting to ask what may be the cause of the large in-plane
ellipticity of the LMC. The prime candidate is distortion by the Milky
Way tidal field. The present-day tidal force on the LMC by the Milky
Way exceeds that from the SMC. Moreover, the Milky Way tidal field is
responsible for other well-known features of the Magellanic system,
such as the Magellanic Stream (see Section~\ref{s:stream}). N-body
simulations have shown that the structure of the LMC can be altered
significantly by the Milky Way tidal force (Weinberg 2000) and the
simulations of Mastropietro et al.~(2004) indeed predict a
considerable in-plane elongation for the LMC. Also, the LMC elongation
in projection on the sky points approximately towards the Galacic
Center and is perpendicular to the Magellanic Stream (van der Marel
2001), as predicted naturally by simulations of tidal effects
(Mastropietro et al.~2004). However, a very detailed data-model
comparison is not possible at the present time. That would require
accurate knowledge of the past history as a function of time of the
LMC orbit, of the disk-plane orientation due to precession and
nutation, and of the LMC-SMC distance. Such knowledge is not available
at the present time.

A consequence of the ellipticity of the LMC disk is that one cannot
expect the streamlines of tracers in the disk to be perfectly
circular, by contrast to what has been assumed in all kinematical
studies to date. The effect of this is probably not large, because the
gravitational potential of a mass distribution is always rounder than
the mass distribution itself. One effect of ellipticity is an apparent
offset between the kinematical line of nodes and the true line of
nodes (Schoenmakers, Franx \& de Zeeuw 1997). This effect is presently
not at observable levels, given that kinematical analysis of carbon
stars (see Section~\ref{s:kinematics}) yields a line-of-nodes position
angle $\Theta$ that is in adequate agreement with geometrical
determinations (see Section~\ref{s:viewing}). However, it should be kept
in mind that, as the sophistication of the studies of LMC kinematics
increases, it might become necessary to account for the effect of
non-circularity on the observed kinematics.

\section{Vertical Structure and Microlensing} 
\label{s:vertical}

\subsection{Scale Height}
\label{ss:height}

The scale height of the LMC disk can be estimated from the observed
line-of-sight velocity dispersion $\sigma$. For carbon stars, the
scatter of the velocity measurements around the best-fitting rotating
disk model (Figure~\ref{f:fit}) yields $\sigma = 20.2 \pm 0.5 \kms$.
The ratio of rotation velocity to velocity dispersion is therefore
$V/\sigma = 2.9 \pm 0.9$. For comparison, the thin disk of the Milky
Way has $V /\sigma \approx 9.8$ and its thick disk has $V/\sigma
\approx 3.9$. In a relative sense one might therefore expect the LMC
disk to be similar, but somewhat thicker than the Milky Way thick
disk. Weinberg (2000) argued from $N$-body simulations that such
considerable thickness could be the result of Milky Way tidal effects
on the LMC.

The radial profile of the velocity dispersion contains information on
the LMC scale height as function of radius. The carbon star velocity
dispersion is close to constant as function of radius, and this is not
what is expected for a disk with a constant scale height. To fit this
behavior one must assume that the scale height increases with radius
in the disk (Alves \& Nelson 2000). This can arise naturally as a
result of tidal forces from the Milky Way, which become relatively
more important (compared to the LMC self-gravity) as one moves to
larger radii. Alves \& Nelson considered an isothermal disk with a
vertical density profile proportional to ${\rm sech}^2 (z/z_0)$, where
$z_0$ can vary with disk radius. Application to the carbon star data
of van der Marel \etal (2002) yields $z_0 = 0.27 \kpc$ at the LMC
center, rising to $z_0 = 1.5 \kpc$ at a radius of $5.5 \kpc$.

The LMC carbon stars are part of the intermediate-age population which
is believed to be fairly representative for the bulk of the mass in
the LMC. In this sense, the results inferred for the carbon star
population are believed to be characteristic for the LMC as a
whole. However, it is certainly not true that all populations have the
same kinematics. As in the Milky Way, younger populations have a
smaller velocity dispersion (and hence a smaller scale height) than
older populations. A summary of measurements for various populations
is given by Gyuk, Dalal \& Griest (2000). The youngest populations
(e.g., supergiants, HII regions, HI gas) have dispersions of only
$\sigma \approx 6 \kms$. Old long-period variables have dispersions
$\sigma \approx 30 \kms$ (Bessell, Freeman \& Wood 1986) and so do the
oldest star clusters (Freeman \etal 1983; Schommer \etal 1992). These
values are considerably below the LMC circular velocity (see
Section~\ref{ss:carbonstars}). This has generally been interpreted to
mean that the LMC does not have an old pressure supported halo similar
to that of the Milky Way.

The first possible evidence for the presence of a pressure supported
halo was presented recently by Minniti \etal (2003). They measured a
dispersion $\sigma \approx 53 \pm 10 \kms$ for a sample of 43 RR Lyrae
stars within $1.5^{\circ}$ from the LMC center. This value is
consistent with what would be expected for a pressure supported halo
in equilibrium in the gravitational potential implied by the circular
velocity (Alves 2004a). The RR Lyrae stars make up $\sim 2$\% of the
visible mass of the LMC, similar to the value for the Milky Way
halo. However, it is surprising that the surface density distribution
of the LMC RR Lyrae stars is well fit by an exponential with the same
scale length as inferred for other tracers known to reside in the disk
(Alves 2004a). This is very different from the situation for the Milky
Way halo, where RR Lyrae stars follow a power-law density
profile. This suggests that maybe the RR Lyrae stars in the LMC did
form in the disk, instead of in a halo. In this scenario they might
simply have attained their large dispersions by a combination of disk
heating and Milky Way tidal forces (Weinberg 2000). To discriminate
between halo and disk origins it will be important to determine
whether the velocity field of the RR Lyrae stars has a rotation
component. This will require observations at larger galactocentric
distances.\looseness=-2

\subsection{Microlensing Optical Depth}
\label{ss:lensing}

One of the most important reasons for seeking to understand the
vertical structure of the LMC is to understand the results from
microlensing surveys. The observed microlensing optical depth towards
the LMC is $\tau_{\rm obs} = 12^{+4}_{-3} \times 10^{-8}$ with an
additional 20--30\% of systematic error (Alcock \etal 2000a). It has
generally been found that equilibrium models for the LMC do not
predict enough LMC self-lensing events to account for the observed
optical depth (Gyuk, Dalal \& Griest 2000; Jetzer, Mancini \&
Scarpetta 2002). For the most favored set of LMC model parameters in
the Gyuk \etal study the predicted self-lensing optical depth is
$\tau_{\rm self} = 2.2 \times 10^{-8}$, a factor of $5.5$ less than
the observed value. To account for the lenses it is therefore
necessary to assume that some $\sim 20$\% of the Milky Way dark halo
is made up of lenses of mass $0.15$--$0.9 \Msun$ (Alcock \etal
2000a). However, it is a mystery what the composition of this lensing
component could be. A large population of old white dwarfs has been
suggested, but this interpretation is not without problems (e.g.,
Fields, Freese, \& Graff 2000; Flynn, Holopainen, \& Holmberg
2003). It is therefore worthwhile to investigate whether the models
that have been used to estimate the LMC self-lensing might have been
oversimplified. This is particularly important since there is evidence
from the observed microlensing events themselves that many of the
lenses may reside in the LMC (Sahu 2003).

To lowest order, the self-lensing optical depth in simple disk models
of the LMC depends exclusively on the observed velocity dispersion and
not separately on either the galaxy mass or scale height (Gould
1995). One way to increase the self-lensing predicted by LMC models is
therefore to assume that a much larger fraction of the LMC mass
resides in high velocity dispersion populations than has previously
been believed.  Salati \etal (1999) showed that the observed optical
depth can be reproduced if one assumes that 70\% of the mass in the
LMC disk consists of objects with $\sigma$ ranging from $25 \kms$ to
$60 \kms$. However, this would seem difficult to explain on the basis
of present data. Although RR Lyrae stars have now been observed to
have a high velocity dispersion, these old stars make up only 2\% of
the visible mass. So their influence on self-lensing predictions is
negligible. A better way to account for the observed optical depth
might be to assume that the vertical structure of the LMC is more
complicated than for normal disk galaxies. The self-lensing optical
depth might then have been considerably underestimated.\looseness=-2

\subsection{Foreground and Background Populations}
\label{ss:foreback}

One way to explain the observed microlensing optical depth is to
assume that there might be stellar populations outside of the main LMC
disk plane. For example, there might be a population of stars in front
of or behind the LMC that was pulled from the Magellanic Clouds due to
Milky Way tidal forces (Zhao 1998) or there might be a non-virialized
shroud of stars at considerable distances above the LMC plane due to
Milky Way tidal heating (Weinberg 2000; Evans \& Kerins 2000). If
microlensing source stars belong to such populations, then this would
yield observable signatures in their characteristics. In particular,
if microlensing source stars are behind the LMC, then they should be
systematically fainter (Zhao, Graff \& Guhathakurta 2000) and redder
(Zhao 1999, 2000) than LMC disk stars. HST/WFPC2 CMDs of fields
surrounding microlensing events show no evidence for this, although
the statistics are insufficient to rule out specific models with
strong confidence (Alcock \etal 2001). Other tests of the spatial
distribution and properties of the LMC microlensing events also do not
(yet) discriminate strongly between different models for the location
and nature of either the lenses or the sources (Alcock \etal 2000a;
Gyuk, Dalal \& Griest 2000; Jetzer, Mancini \& Scarpetta 2002). This
is primarily because only 13--17 LMC microlensing events are known. So
to assess whether the LMC has out-of-plane structures it is best to
focus on other types of tracers that might yield better statistics.

Zaritsky \& Lin (1997) studied the optical CMD of a field $\sim
2^{\circ}$ north-west of the LMC center and found a vertical extension
at the bright end of the Red Clump. They suggested that this feature
is due to stars 15--$17 \kpc$ above the LMC disk. However, this
interpretation has been challenged for many different reasons (e.g.,
Bennett 1998; Gould 1998, 1999). Most seriously, Beaulieu \& Sackett
(1998) pointed out that a Vertical Red Clump (VRC) extension is
naturally expected from stellar evolution due to young helium core
burning stars. The same feature is seen in the Fornax and Sextans A
dwarfs (Gallart 1998). It remains somewhat open to discussion whether
the number of stars in the LMC VRC is larger than expected given our
understanding of the LMC star formation history, so a foreground
population is not strictly ruled out (Zaritsky \etal 1999). However,
it certainly doesn't appear to be a favored interpretation of the
data. Also, the kinematics of the VRC stars is indistinguishable from
that of LMC Red Clump stars (Ibata, Lewis \& Beaulieu 1998). More
recently, Zhao \etal (2003) performed a detailed radial velocity
survey of 1300 stars of various types within $\sim 2^{\circ}$ from the
LMC center. They found no evidence for stars with kinematics that
differ significantly from that of the main LMC disk. This rules out a
significant foreground or background population of stars that reside
in a tidal stream that is seen superposed onto the LMC by chance (Zhao
1999). Any foreground or background population must be physically
associated with the LMC, and share its kinematics.  Sub-populations
within the LMC with subtly different kinematics have been suggested
(Graff \etal 2000), but only at low statistical significance.

Other arguments for stars at large distances from the LMC disk also
have not been convincing. Kunkel \etal (1997b) argued on the basis of
carbon star velocities that the LMC has a polar ring. However, the
carbon star velocity field shown in Figure~\ref{f:fit} seems to be
well fit by a single rotation disk model. It is possible that the
Kunkel \etal study was affected by the use of an LMC transverse
velocity value of only $240 \kms$, which is considerably below the
value indicated by presently available proper motion data ($406 \pm 44
\kms$; see Section~\ref{s:orbit}). Weinberg \& Nikolaev (2001) found a
tail of relatively faint stars in luminosity functions of AGB stars
selected by $J-K$ color from 2MASS data. They suggested that this is
not due to dust extinction but might indicate stars behind the LMC. On
the other hand, the distribution of these stars on the sky (van der
Marel, unpublished) bears a strong resemblance to the far infrared
IRAS map of LMC dust emission (Schwering 1989). This casts doubt on
the interpretation that the brightnesses of these stars have not been
affected by dust. Weinberg \& Nikolaev (2001) also found a slightly
non-Gaussian tail in their AGB star luminosity functions towards
brighter magnitudes. However, this need not indicate stars in the
foreground, given that there is no a priori reason why the AGB star
luminosity function would have to be Gaussian. In another study,
Alcock \etal (1997) found no evidence for unexpected numbers of RR
Lyrae stars at distances beyond $\sim 15 \kpc$ from the LMC disk
plane.

\subsection{Non-planar Structure in the LMC Disk}
\label{ss:nonplanar}

The studies discussed in Sections~\ref{s:kinematics}
and~\ref{s:viewing} indicate that the overall structure of the LMC is
that of a (thick) disk. From Section~\ref{ss:foreback} it follows that
there is no strong evidence for unexpected material far from the disk
plane. However, this does not mean that there may not be non-planar
structures in the disk itself. For example, it was already discussed
in Section~\ref{s:viewing} that there might be warps and twists in the
disk plane.

Early evidence for non-planar material came from HI observations. One
of the most prominent optical features of the LMC is the star forming
30 Doradus complex, located just north of the eastern tip of the
bar. This region is a very strong source of UV radiation, yet the HI
gas disk of the LMC does not show a void in this part of the sky. This
indicates that the 30 Doradus complex cannot be in the plane of the
LMC disk, but must be at least $250$--$400 \kpc$ away from it (Luks \&
Rohlfs 1992). HI channel maps show that there is a separate HI
component, called the ``L-component'', that is distinct from the main
LMC disk. It has lower line-of-sight velocities than the main disk by
$20$--$30 \kms$, contains some 19\% of the HI gas in the LMC, and does
not extend beyond $2^{\circ}$--$3^{\circ}$ from the LMC center (Luks
\& Rohlfs 1992). Absorption studies indicate that this component is
behind the LMC disk (Dickey \etal 1994). The 30 Doradus complex is
spatially located at the center of the L-component. It is probably
directly associated with it, given that the L-component shows a hole
of HI emission at the 30 Doradus position as expected from ionization.
These results suggest that in the central few degrees of the LMC an
important fraction of the gas and stars may not reside in the main
disk.

%%% FIGURE 10 %%%

\begin{figure}
\null\bigskip
\epsfxsize=0.6\hsize
\centerline{\epsfbox{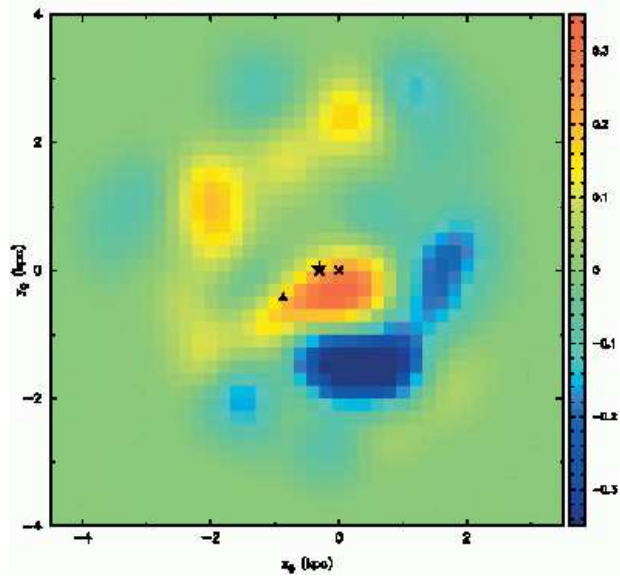}}
\caption{Map of the average vertical distances of Cepheids (in kpc)
from a best-fitting plane solution as function of in-plane coordinates
$(x_0,y_0)$ from Nikolaev \etal (2004). The orientation of the bar in
this representation is similar as in Figure~\ref{f:maps}. Negative
(positive) distances denote material behind (in front of) the fitted
plane. The cross indicates the HI rotation center according to Kim
\etal (1998), the star shows the geometric center of the Cepheid
sample, and the triangle gives the center of the outer carbon star
isophotes from van der Marel \& Cioni (2001). A color version of the
image is available in the Nikolaev \etal paper.\label{f:cepheids}}
\end{figure}

%%% END OF FIGURE 10 %%%

It has now proven possible to test some of these ideas more directly
by using large stellar databases. Section~\ref{s:viewing} already
mentioned the study of more than 2000 Cepheids by Nikolaev \etal
(2004). They obtained accurate reddening-corrected (relative)
distances to each of the Cepheids. The distance residuals with respect
to the best-fitting plane do not follow a random Gaussian
distribution. Instead they show considerable structure, as shown in
Figure~\ref{f:cepheids}. The two-dimensional distribution of the
residuals on the sky was interpreted as a result of two effects,
namely a symmetric warp in the disk, and the fact that the bar is
located $\sim 0.5 \kpc$ in front of the main disk. Interestingly, an
offset between the LMC bar and disk had previously been suggested by
Zhao \& Evans (2000) as an explanation for the observed LMC
microlensing optical depth. If the disk and the bar of the LMC are not
dynamically connected, then this may also explain why in projection
the LMC bar appears offset from the center of the outer isophotes.

The variation of the reddening-corrected Red Clump magnitude over the
face of the LMC has also been used to study vertical structures in the
LMC disk. Olsen \& Salyk observed 50 randomly selected fields in the
central $4^{\circ}$ of the LMC at CTIO. They found that fields between
$2^{\circ}$--$4^{\circ}$ south-west of the LMC center are $0.1$ mag
brighter than expected from the best plane fit. They interpreted this
to indicate that stars in these fields lie some $2 \kpc$ above the LMC
disk, and argued for warps and twists in the LMC disk
plane. Subramaniam (2003) used Red Clump magnitudes determined from
stars in the OGLE database to address the same issue. This yielded a
map of the Red Clump magnitude along the length and width of the LMC
bar. This map shows considerable structure and clearly cannot be fit
as a single plane. Subramaniam (2004) suggested that the residuals can
be interpreted as the result of a misaligned secondary bar inside the
primary bar.

Eclipsing binaries have also provided interesting information on this
subject. These binaries can be modeled in detail to yield a fairly
accurate distance. The distances of sources studied so far seem to
indicate a slightly lower LMC distance modulus of $m-M \approx 18.4$
(Ribas \etal 2002) than has been inferred from other tracers (see
Section~\ref{s:orbit}). However, it is possible to obtain somewhat
higher values with a slightly different analysis (Groenewegen \&
Salaris 2001; Clausen \etal 2003) so this is no great cause for
concern (Alves 2004b). What is interesting though is that four
eclipsing binaries analyzed by the same team with the same method show
a considerable spread in distance. This has been interpreted to mean
than one of the binaries lies $\sim 3 \kpc$ behind the LMC disk plane
(Ribas \etal 2002) and that another one lies $\sim 4 \kpc$ in front of
it (Fitzpatrick \etal 2003).

The above studies indicate that there is considerable and complicated
vertical structure in the central few degrees of the LMC
disk. However, this is a rapidly developing field, and many important
questions remain open. In particular, it is unclear whether the
features reported in the various studies are actually the same or
not. Qualitative comparison is not straightforward. Different authors
study different areas of the LMC, they plot residuals with respect to
different planes, and they present their results in figures that plot
different types of quantities. However, cursory inspection of the
various papers shows very few features that are obviously in common
between the studies. Quantitative comparison is therefore
needed. Direct comparison to the results for the outer parts of the
LMC, where there is little evidence for extra-planar structures (van
der Marel \& Cioni 2001), is also important. If discrepancies emerge
from such comparisons, then this can mean two things. Either some
studies are in error (e.g., due to use of inaccurate dust corrections,
or by incorrect interpretation of stellar evolutionary variations as
distance variations) or different tracers do not trace the same
structure. The latter might well the case. Cepheids are young stars
with ages less than a few times $10^8$ years whereas stars on the RGB
and AGB are typically older than 1 Gyr.  Since the structure of the
LMC and its gaseous component vary with time as a result of tidal
interactions with the Milky Way and the SMC, one wouldn't necessarily
expect stars formed at different epochs to trace identical structures.

Another open question is what the physical and dynamical
interpretation is of the extra-planar structures that are being
detected.  Many authors have used the term ``warp''. However, the
residuals that have been reported do not much resemble the smoothly
varying residuals that are expected from a single plane with a global
warp. The structures appear to be both different and more complicated
than a single warp. If a decomposition into different components is
attempted, then such a decomposition must use entities that can be
understood dynamically (disk, bar, bulge, warp, etc.) and that make
sense in the context of the overall evolutionary history of the
Magellanic System. Components with holes or sharp edges (e.g.,
Subramaniam 2004) are not particularly realistic from a dynamical
viewpoint. Phase mixing of stellar orbits quickly removes such sharp
discontinuities. Also, if the vertical structures detected in the
inner region of the LMC disk are due to components that are not
connected to the main disk plane, then why does the projected image of
the LMC (e.g., Figure~\ref{f:maps}) look so smooth? And why is there
so little evidence from stellar kinematics for components with
decoupled kinematics (e.g., Figure~\ref{f:fit}; Zhao \etal 2003)?
Clearly, many questions remain to be answered before we can come to a
full understanding of the vertical structure of the LMC.

\section{Concluding Remarks}
\label{s:conc}

The structure and kinematics of the LMC continue to be active areas of
research. As outlined in this review, much progress has been made
recently. Improved datasets have played a key role in this, most
notably the advent of large stellar datasets of magnitudes in many
bands, lightcurves, and line-of-sight kinematics, and also the
availability of sensitive HI observations over large areas. As a
result we now have a fairly good understanding of the LMC morphology
and kinematics. The proper motion of the LMC is reasonably well
measured and the global properties of the LMC orbit around the Milky
Way are understood. The angles that determine how we view the LMC are
now known much more accurately than before and this has led to the
realization that the LMC is quite elliptical in its disk plane. We are
starting to delineate the vertical structure of the LMC and are
finding complexities that were not previously expected.

Despite the excellent progress, many questions on LMC structure still
remain open. Why is the bar offset from the center of the outer
isophotes of the LMC? Why is the dynamical center of the HI offset
from the center of the bar, from the center of the outer isophotes,
and from the dynamical center of the carbon stars? Why do studies of
the inner and outer regions of the LMC yield differences in
line-of-nodes position angle of up to $30^{\circ}$? Does the LMC have
a pressure supported halo? Are there populations of stars at large
distances from the LMC plane? What is the origin and dynamical nature
of the non-planar structures detected in the inner regions of the LMC?
Do different tracers outline the same non-planar structures?  

It might be necessary to answer all of these open questions before we
can convince ourselves that the optical depth for LMC self-lensing has
been correctly estimated. This seems to be the most critical step in
establishing whether or not the Milky Way halo contains hitherto
unknown compact lensing objects (MACHOs). The open questions about LMC
structure are important also in their own right. The tidal interaction
between the Magellanic Clouds and the Milky Way provides one of our
best laboratories for studying the processes of tidal disruption and
hierarchical merging by which all galaxies are believed to grow. A
better understanding of LMC structure may also provide new insight
into the origin of the Magellanic Stream, which continues to be
debated. And with improved proper motion measurements of the
Magellanic Clouds, the Stream may become a unique tool to constrain
the shape and radial density distribution of the Milky Way halo at
radii inaccessible using other tracers.

%%%%%%%%%%%%%%%%%%%%
% References
%%%%%%%%%%%%%%%%%%%%

\end{document}